\documentclass[fleqn,usenatbib]{mnras}

\usepackage{newtxtext,newtxmath}
\usepackage[T1]{fontenc}
\usepackage{ae,aecompl}

\usepackage{graphicx}	% Including figure files
\usepackage{amsmath}	% Advanced maths commands
\usepackage{gensymb}
\usepackage[flushleft]{threeparttable}
\usepackage[usenames, dvipsnames]{color}

\title[An extreme HB star in the LMC]{An Extreme Amplitude, Massive Heartbeat System in the LMC Characterized Using ASAS-SN and TESS}

\author[T. Jayasinghe et al.]{T. Jayasinghe$^{1,2}$\thanks{E-mail: jayasinghearachchilage.1@osu.edu},
K. Z. Stanek$^{1,2}$,
C. S. Kochanek$^{1,2}$,
Todd A. Thompson$^{1,2,3}$,
\newauthor 
B. J. Shappee$^{4}$,
M. Fausnaugh$^{5}$
\newauthor 
\\
$^{1}$Department of Astronomy, The Ohio State University, 140 West 18th Avenue, Columbus, OH 43210, USA\\
$^{2}$Center for Cosmology and Astroparticle Physics, The Ohio State University, 191 W. Woodruff Avenue, Columbus, OH 43210, USA\\
$^{3}$Institute for Advanced Study, Princeton, NJ, 08540, USA\\
$^{4}$Institute for Astronomy, University of Hawaii, 2680 Woodlawn Drive, Honolulu, HI 96822, USA\\
$^{5}$MIT Kavli Institute for Astrophysics and Space Research, 77 Massachusetts Avenue, 37-241, Cambridge, MA 02139, USA\\
}

\date{Accepted XXX. Received YYY; in original form ZZZ}

\pubyear{2018}
 
\begin{document}
\label{firstpage}
\pagerange{\pageref{firstpage}--\pageref{lastpage}}
\maketitle

\begin{abstract}
Using ASAS-SN data, we find that the bright ($V{\sim}13.5$ mag) variable star MACHO 80.7443.1718 (ASASSN-V J052624.38-684705.6) is the most extreme heartbeat star yet discovered. This massive binary, consisting of at least one early B-type star, has an orbital period of $P_{\rm ASAS-SN}=32.83627\pm0.00846\,{\rm d},$  and is located towards the LH58 OB complex in the LMC. Both the ASAS-SN and TESS light curves show extreme brightness variations of ${\sim}40\%$ at periastron and variations of ${\sim}10\%$ due to tidally excited oscillations outside periastron. We fit an analytical model of the variability caused by the tidal distortions at pericenter to find orbital parameters of $\omega=-61.4\degree$, $i=44.8\degree$ and $e=0.566$.  We also present a frequency analysis to identify the pulsation frequencies corresponding to the tidally excited oscillations.
\end{abstract}


%%%%%%%%%%%%%%%%%%%%%%%%%%%%%%%%%%%%%%%%%%%%%%%%%%

%%%%%%%%%%%%%%%%% BODY OF PAPER %%%%%%%%%%%%%%%%%%

\section{Introduction}

Heartbeat stars are short period ($P\lesssim1$\,yr), eccentric ($e\gtrsim0.3$) binaries where oscillations are excited by the tidal forcing at each periastron passage. Heartbeat stars were first discovered in data from the \textit{Kepler} space telescope \citep{2012ApJ...753...86T}. The prototypical heartbeat star, KOI-54, has been extensively studied and characterized (see for e.g., \citealt{2011ApJS..197....4W,2012MNRAS.420.3126F,2012MNRAS.421..983B} and references therein), and \textit{Kepler} \citep{2016AJ....151...68K} has now identified over 170 heartbeat stars.

The light curves of heartbeat stars are defined by oscillations outside of periastron combined with a brief, high amplitude ellipsoidal variation at periastron that gives rise to a unique ``heartbeat'' signature resembling the normal sinus rhythm of an electrocardiogram. The light curves of these systems are dominated by the effects of tidal distortion, reflection and Doppler beaming close to periastron \citep{2017MNRAS.472.1538F}. Heartbeat stars continue to oscillate throughout their orbit due to tidally excited stellar oscillations. The variability amplitude of most heartbeat stars is very small ($\lesssim 1$ mmag; \citealt{2016AJ....151...68K,2018MNRAS.473.5165H}).

The tidally excited oscillations (TEOs) occur at exact integer multiples of the orbital frequency \citep{2017MNRAS.472.1538F}. TEOs were first discovered in the eccentric binary system HD 174884 \citep{2009A&A...508.1375M} and later confirmed in KOI 54 and several other systems \citep{2016ApJ...829...34S}. The largest amplitude TEOs are driven by resonances between harmonics of the orbital frequency and the normal mode frequencies of the star. Both the amplitudes and phases of TEOs can be predicted from linear theory \citep{2017MNRAS.472.1538F}.

The vast majority of the heartbeat stars that have been discovered are relatively low-mass A and F type stars. However, the heartbeat phenomenon extends to more massive OB stars as well. $\iota$ Ori is the most massive heartbeat star system yet discovered and it consists of a O9 III primary and a B1 III-IV companion \citep{2017MNRAS.467.2494P}. The dearth of massive heartbeat stars is likely an observational bias because massive stars are rare and \textit{Kepler} observed only a small fraction of the sky, mostly lying off the Galactic plane. Ground-based surveys cover most or all of the sky but find it challenging to detect the low variability amplitudes of typical heartbeat stars ($\Delta L/L\sim 10^{-3}$; \citealp{2017MNRAS.472.1538F}).

The All-Sky Automated Survey for SuperNovae (ASAS-SN, \citealt{2014ApJ...788...48S, 2017PASP..129j4502K}) has been monitoring the entire visible sky for several years to a depth of $V\lesssim17$ mag with a cadence of $2-3$ days using two units in Chile and Hawaii, each with 4 telescopes. As of the end of 2018, ASAS-SN uses 20 telescopes to observe the entire sky daily, but all current observations are taken with a $g$-band filter. We have written a series of papers studying variable stars using ASAS-SN data. In Paper I \citep{2018MNRAS.477.3145J}, we reported ${\sim}66,000$ new variables that were discovered during the search for supernovae. In Paper II \citep{2018arXiv180907329J}, we homogeneously analyzed ${\sim} 412,000$ known variables from the VSX catalog, and developed a robust variability classifier utilizing the ASAS-SN V-band light curves and data from external catalogues. In Paper III \citep{2019MNRAS.485..961J}, we conducted a variability search towards the Southern Ecliptic pole in order to overlap with the The Transiting Exoplanet Survey Satellite (TESS; \citealt{2015JATIS...1a4003R}) continuous viewing zone (CVZ).  We identified ${\sim}11,700$ variables, of which ${\sim}7,000$ were new discoveries.

TESS is currently conducting science operations by monitoring (eventually) most of the sky with a baseline of at least 27 days. Sources closer to the TESS CVZ will be observed for a substantially longer period, approaching one year at the ecliptic poles. TESS full-frame images (FFIs), sampled at a cadence of 30 min, are made publicly available, allowing for the study of short time scale variability across most of the sky.

Here we discuss the identification of the most extreme amplitude heartbeat star yet discovered, MACHO 80.7443.1718 (ASASSN-V J052624.38-684705.6, TIC 373840312), using both ASAS-SN and TESS photometry. The MACHO survey reported that the source was a variable, but classified it as an eclipsing binary \citep{1997ApJ...486..697A}. MACHO 80.7443.1718 was first identified as a likely heartbeat star during our ASAS-SN variability search. We discuss archival data and the ASAS-SN and TESS observations in section 2. In section 3, we fit an analytical model for the tidal distortions to estimate several orbital parameters of the binary system. In section 4, we discuss our SED fits to this source and the physical implications of these fits. In section 5, we identify tidally excited oscillations through a periodogram analysis and present a summary of our results in section 6. 

\section{Archival, ASAS-SN and TESS data for MACHO 80.7443.1718}

The source MACHO 80.7443.1718 was identified as a variable by the MACHO survey \citep{1997ApJ...486..697A}, who classified it as a generic eclipsing binary. Two values for the orbital period, corresponding to the ``red'' and ``blue'' bandpasses, were derived using their data: $P_{\rm red}=32.83108\;$d, $P_{\rm blue}=32.83397\;$d. 

MACHO 80.7443.1718 is a blue source with $U-B=-0.84$ mag, $B-V=0.11$\,mag with estimated values for the photometric temperature $\log{(\rm T_{\rm *}/\textrm{K})}=4.6$ and bolometric magnitude $M_{\rm bol}=-9.1$ \citep{2002ApJS..141...81M}. This source is part of the LH58 OB association in the Large Magellanic Cloud (LMC), northwest of 30 Doradus. An archival spectrum classified it as a B0.5 Ib/II \citep{1994AJ....108.1256G}, evolved blue star. Based on this information, this source is likely to have a mass $M\gtrsim10 M_\odot$ \citep{2014A&A...566A...7N}.

The Gaia DR2 \citep{2018arXiv180409365G} counterpart is {\verb"source_id=4658489067332871552"}. Its nominal DR2 parallax is negative ($\pi=-0.0586 \pm0.0238 \,\rm mas$) and the proper motion is $\mu_{\alpha}=1.59\pm0.04 \, \rm mas\;yr^{-1}$, $\mu_{\delta}=0.66\pm0.05 \, \rm mas\;yr^{-1}$. There might be evidence of problems with the astrometric solution, as the $\chi^2$ of the fit is high and the excess astrometric noise is $0.14$~mas.  On the other hand, the renormalized unit weight error  (RUWE; \citealt{2018A&A...616A...2L}) of the source is 0.96 and  \cite{2018A&A...616A...2L} argue that Gaia DR2 astrometric solutions are accurate if they have a ${\rm RUWE}<1.4$.

To evaluate the proper motions, we examined 56 Gaia DR2 stars with $G<15$~mag within $5'$ of MACHO~88.7443.1718.  These 56 stars had medians ($1\sigma$ range)
of $-0.024$~mas ($-0.126<\pi<0.020$), $1.670\;\rm mas\;yr^{-1}$ ($0.540 < \mu_\alpha <1.814$)
and $0.697\;\rm mas\;yr^{-1}$ ($0.125 < \mu_\delta < 0.852$) for their parallax and proper motions.  Hence the parallax and proper motions of MACHO~88.7443.1718 are typical of the local population of luminous stars.  If we use the median proper motions to define a local standard of rest, the relative motion of MACHO~88.7443.1718 is $0.088\;\rm mas\;yr^{-1}$, or roughly $21\;\rm km\;s^{-1}$ at a distance of $50$~kpc.  The median motion of the nearby stars relative to this standard of rest is $0.215\;\rm mas\;yr^{-1}$ or roughly $50\;\rm km\;s^{-1}$.

ASAS-SN V-band observations were made by the ``Cassius" (CTIO, Chile) quadruple telescope between 2013 and 2018. Each camera has a field of view of 4.5 deg$^2$, the pixel scale is 8\farcs0 and the FWHM is ${\sim}2$ pixels. The light curves for this source was extracted as described in \citet{2017PASP..129j4502K} using aperture photometry with a 2 pixel radius aperture. The AAVSO Photometric All-Sky Survey 
(APASS; \citealt{2015AAS...22533616H}) DR9 catalog was used for absolute photometric calibration. The ASAS-SN V-band light curve has a time baseline of ${\sim}1669$ d. 
%32.83627322885312 error 0.00846
We derived possible periods for this source following the procedure described in \citet{2018MNRAS.477.3145J,2018arXiv180907329J}. The \verb"astrobase" implementation \citep{astrob} of the Generalized Lomb-Scargle (GLS, \citealt{2009A&A...496..577Z,1982ApJ...263..835S}), the Multi-Harmonic Analysis Of Variance (MHAOV, \citealt{1996ApJ...460L.107S}), and the Box Least Squares (BLS, \citealt{2002A&A...391..369K}) periodograms were used to search for periodicity in these light curves. We calculated the Lafler-Kinmann \citep{1965ApJS...11..216L,2002A&A...386..763C} string length statistic $T(\phi|P)$  on the phased light curve for each period using the definition
\begin{equation}
    T(\phi|P)=\frac{\sum_{i=1}^{\rm N} (m_{i+1}-m_i)^2}{\sum_{i=1}^{\rm N} (m_{i}-\overline m)^2}\times \frac{(N-1)}{2N}
	\label{eq:tt}
\end{equation} from \citet{2002A&A...386..763C}, where the $m_i$ are the magnitudes sorted by phase and $\overline m$ is the mean magnitude. The best period is the period with the smallest $T(\phi | P)$, which in this case is the best BLS period. 
$$P_{\rm ASAS-SN}=32.83627\pm0.00846\,{\rm d},$$ 
which agrees with the MACHO periods to within ${\sim}0.02$\%. The ASAS-SN light curve hence contains ${\sim}50$ orbits of this system. The error in the period was estimated using the spacing in the BLS frequency grid ($\Delta f=7.84\times 10^{-6} \rm day^{-1}$), where $\Delta P={\Delta f}/{f^2}=0.00846 \, \rm d.$

%the \verb"Period04" software package \citep{2005CoAst.146...53L}.

MACHO 80.7443.1718 lies in the Southern TESS CVZ which allowed us to extract TESS light curves for both Sectors 1 and 2.  Due to the large pixel size of TESS (21") and the crowded region surrounding MACHO 80.7443.171, we used image subtraction \citep{1998ApJ...503..325A,2000A&AS..144..363A} on the full frame images (FFIs) from the first TESS data release to produce high fidelity light curves. The 27 day baseline for TESS observations in each sector is insufficient to obtain a full orbit for this source, but the final TESS light curve with data from all the sectors in the south should sample the variability of this source very well. The difference light curve was normalized to match
the ASAS-SN $V$-band light curve.

\begin{figure*}
	\includegraphics[height=0.95\textheight]{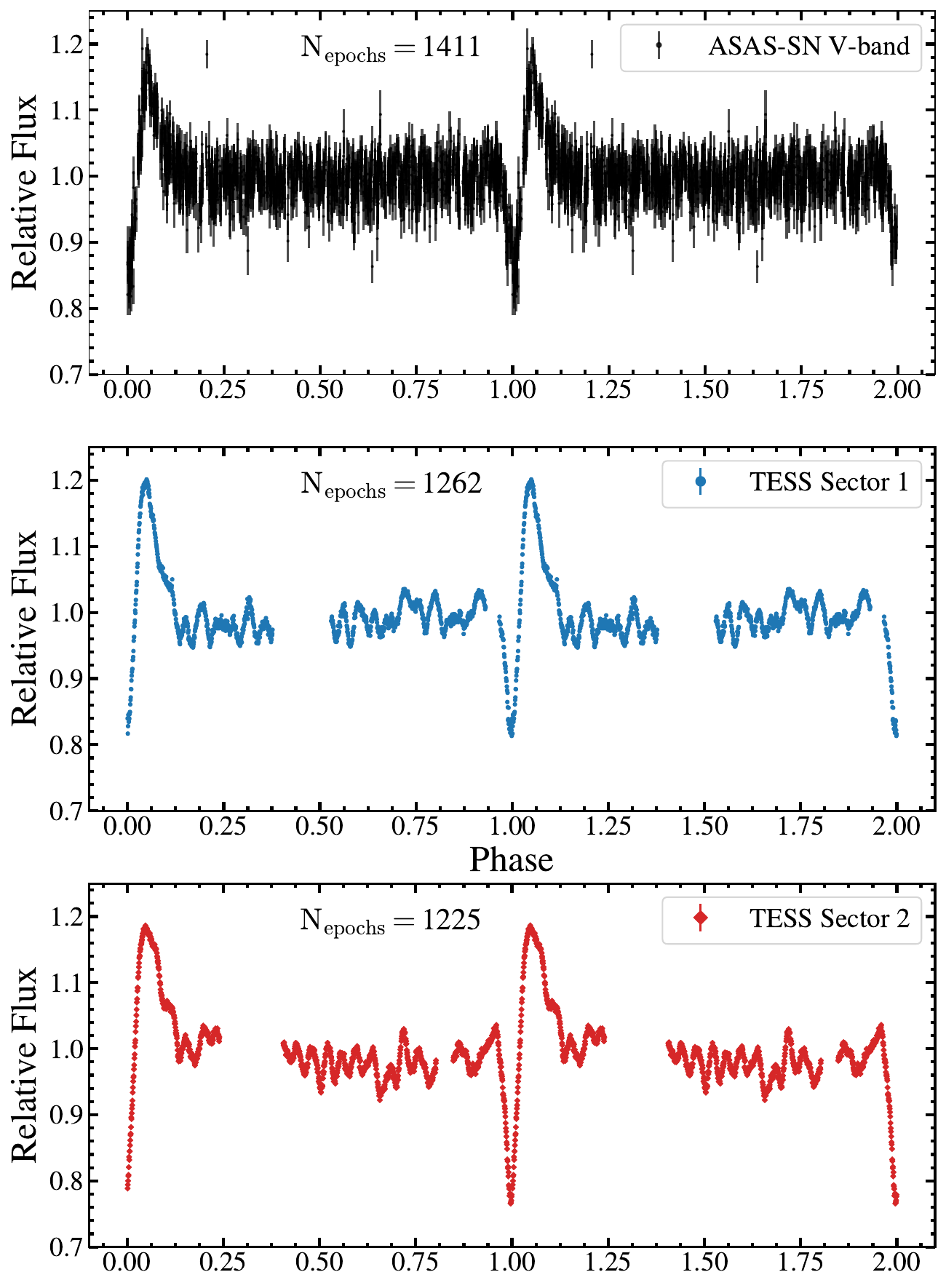}
    \caption{The phased ASAS-SN (top), TESS sector 1 (middle), and TESS Sector 2 (bottom) light curves for the source MACHO 80.7443.1718.}
    \label{fig:fig1}
\end{figure*}

The ephemeris for the minimum of the ellipsoidal variation in ASAS-SN is \begin{equation}
    \rm EphemI= BJD \, 2458143.698310 +32.83627 \times E\,,
	\label{eq:ephemasassn}
\end{equation}whereas the ephemeris for the minimum of the ellipsoidal variation in TESS is
\begin{equation}
    \rm EphemII= BJD \, 2458373.61518 +32.83627 \times E\,,
	\label{eq:ephemtess}
\end{equation}
where the epoch $\rm E$ is the number of orbits since the time of minimum.

The phased ASAS-SN and TESS Sector 1+2 light curves are shown in Figure \ref{fig:fig1}. The uncertainties in relative flux are virtually indistinguishable from the points in the TESS light curves. The TEOs are clearly visible in the TESS light curves (red and green points), but are less distinguishable in the ASAS-SN light curve (blue points).

\begin{figure}
	\includegraphics[width=0.5\textwidth]{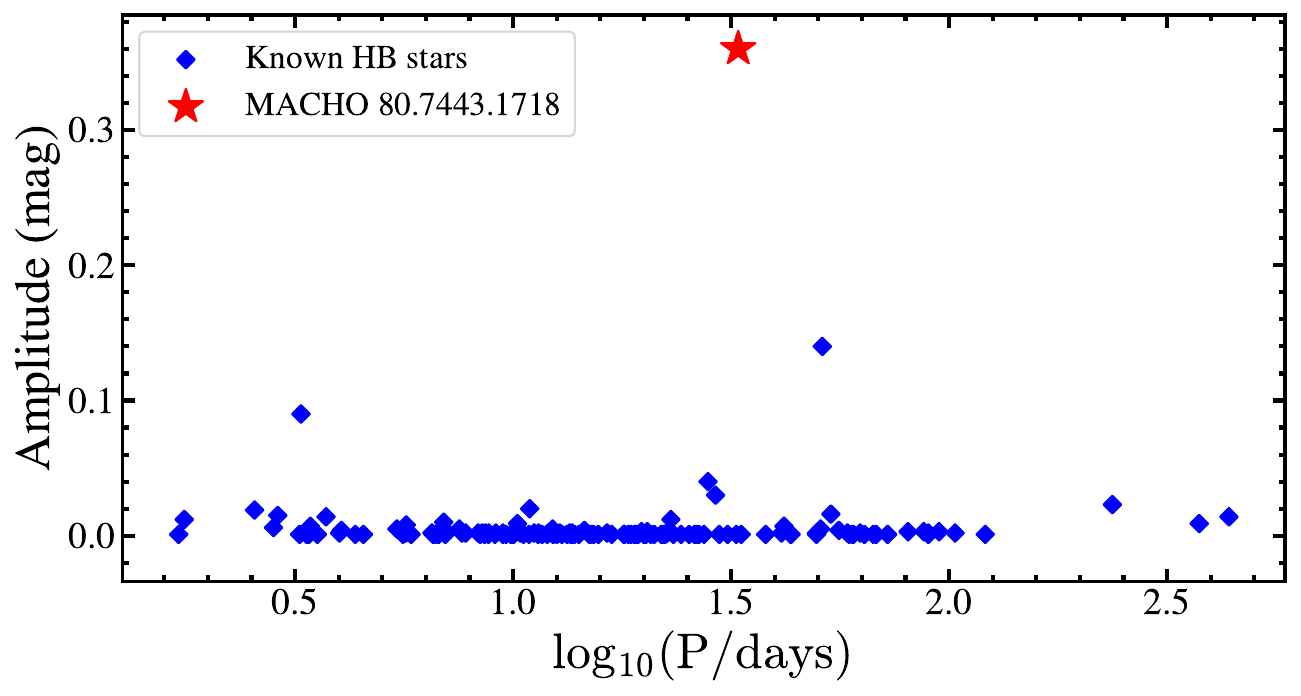}
    \caption{The period-amplitude diagram for the sample of heartbeat stars.  Heartbeat stars  listed in the VSX database \citep{2006SASS...25...47W} are indicated by blue diamonds. MACHO 80.7443.1718 (this work) is shown with a red star. }
    \label{fig:fig2}
\end{figure}

The peak-to-peak flux variations at periastron of ${\sim}40\%$ (${\sim}0.36$ mag) are the largest 
observed for a heartbeat system.  This is 
illustrated in Figure \ref{fig:fig2}, where we compare the period and amplitude of MACHO~80.7443.1718 to those of the heartbeat stars in the VSX catalog \citep{2006SASS...25...47W}.
 The flux variations outside of periastron are also extreme (${\sim}0.1$\,mag).  \citet{2017MNRAS.472.1538F} notes that very large amplitude TEOs are unlikely to arise from a chance resonance and are more likely to stem from a resonantly locked mode. 

\section{Modelling the eccentric ellipsoidal variations}
\citet{1995ApJ...449..294K} developed an analytical model (their Equation 44) for the flux variations produced by the tidal distortions produced by eccentric binaries at periastron. \citet{2012ApJ...753...86T} successfully applied this model to fit the light curves of the heartbeat stars observed by \textit{Kepler}. We fit the fractional flux $\delta F/F$ of the ASAS-SN and TESS light curves following \citet{2012ApJ...753...86T}. The fit contains six parameters: the amplitude scaling factor, $S$, a fractional flux offset, $C$, the true anomaly, $\phi(t)$, the angle of periastron, $\omega$, the orbital inclination, $i$ and the eccentricity, $e$, 
\begin{equation}
    \frac{\delta F}{F}=S \frac{1-3\,\sin^2(i)\sin^2(\phi(t)-\omega)}{(R(t)/a)^3}+C\, ,
	\label{eq:frac}
\end{equation}
where
\begin{equation}
    \frac{R(t)}{a}=1-e\cos(E)\, ,
	\label{eq:rt}
\end{equation}
\begin{equation}
    \phi(t)=2\,\arctan\left(\sqrt{\frac{1+e}{1-e}}\,\tan\left(\frac{E}{2}\right)\right),
	\label{eq:tan}
\end{equation}
and the eccentric anomaly ($E$) is derived from solving Kepler's transcendental equation (e.g., \citealt{2010exop.book...15M}). We performed a trial fit through the Levenberg-Marquardt chi-square minimization routine in \verb"scikit-learn" 
\citep{2012arXiv1201.0490P}. The parameters from the trial fit were then used to initialize a Monte Carlo Markov Chain sampler (MCMC) with 100 walkers, and was then run for 5000 iterations. We used the MCMC implementation through \verb"emcee" \citep{2013PASP..125..306F}. The errors in the parameters were derived from the MCMC chains.

\begin{figure*}
	\includegraphics[width=\textwidth]{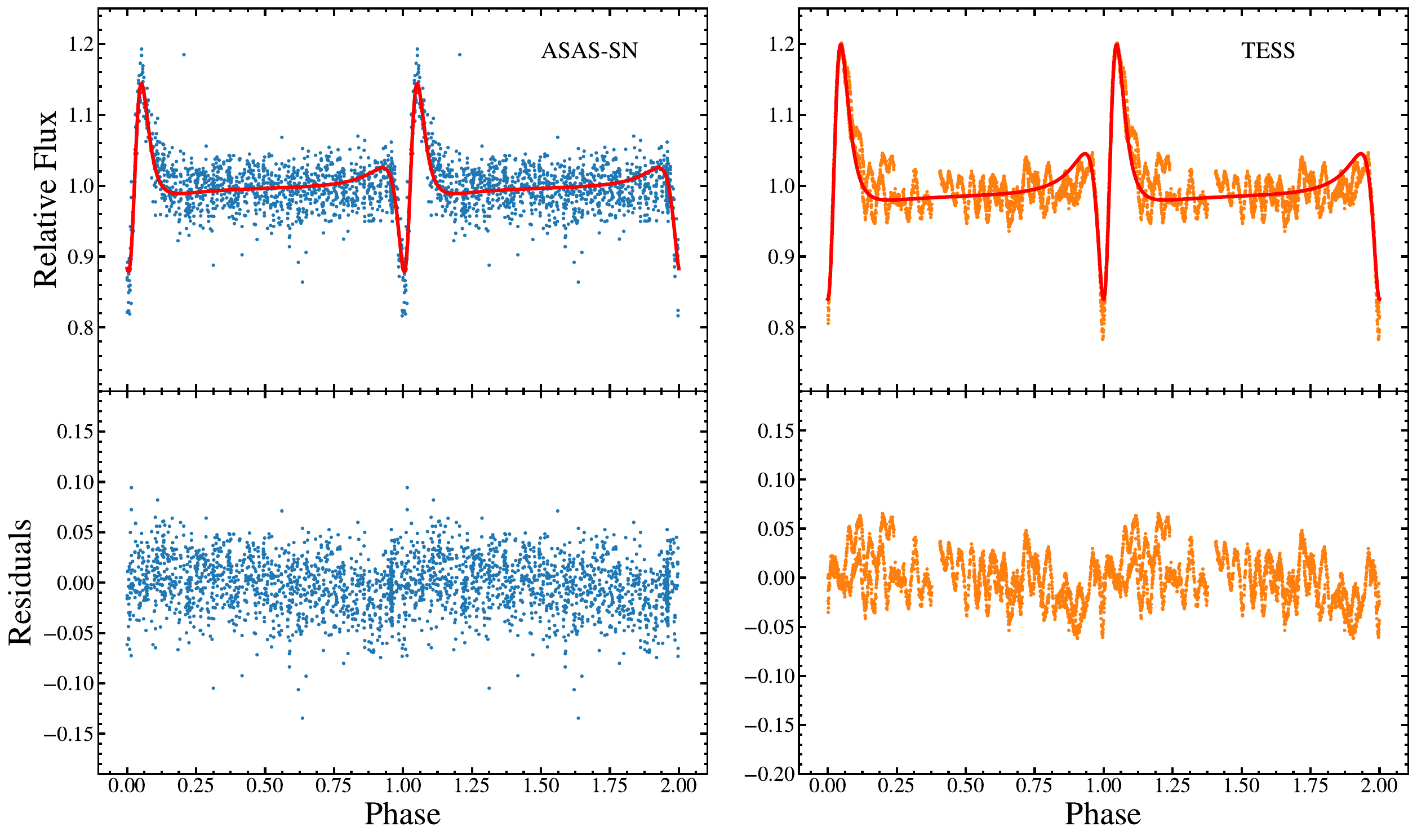}
    \caption{The phased ASAS-SN and TESS light curves (top) and residuals (bottom) for the source MACHO 80.7443.1718 after fitting with the \citet{1995ApJ...449..294K} model for eccentric tidal distortions. Uncertainties in flux are not shown for clarity. The best-fit models are shown in red.}
    \label{fig:fig3}
\end{figure*}

\begin{table}
	\centering
	\caption{Best fit parameters for MACHO 80.7443.1718}
	\label{tab:fits}
\begin{tabular}{rrrr}
		\hline
		 & Description & ASAS-SN & TESS\\
		\hline
		$\omega$ & Angle of periastron &  $-54.50\degree \pm 1.15\degree$ &  $-61.35\degree \pm 0.02\degree$\\
		$i$ & Orbital inclination &  $48.01\degree \pm 0.42\degree $ & $44.77\degree \pm 0.01\degree$\\
		$e$ & Orbital Eccentricity & $0.566\pm 0.004$ & $0.565\pm0.002$ \\
\hline
\end{tabular}
\end{table}

These fits do not capture the depth of the minimum or the height of the maximum completely, suggesting that this model is an incomplete description of the light curve. 
Formally, the estimates for $\omega$ and $i$ from fitting the ASAS-SN and TESS light curves differ by $3\sigma$, but the actual differences of $ < 7^\circ$ are remarkably small given the very different characters of the two light curves. The eccentricities derived from the two fits are consistent to within $1\sigma$. While this model does not account for effects such as irradiation and Doppler boosting, it is a good approximation of the tidal distortions during the orbit and can be used to estimate the orbital parameters of this system without requiring further knowledge about the properties of the stars in the system. The best-fit models for the ASAS-SN and combined Sector 1+2 TESS data for MACHO 80.7443.1718 are shown in the top panel of Figure \ref{fig:fig3} with the solid red lines. The best fit parameters are summarized in Table \ref{tab:fits}. 

We attempted fitting a standard eclipsing binary model including irradiation and reflection effects to the ASAS-SN V-band light curve using \verb"PHOEBE 2.1" \citep{2016ApJS..227...29P,2018ApJS..237...26H} but were unable to replicate the observed variability amplitude.  This is not very surprising because the high amplitude pulses are almost certainly due to transient dynamical tides that are not included in normal eclipsing binary modeling codes. Unfortunately,  there are no public codes for modeling heartbeat stars which include these effects. We have put off attempts at more detailed models until we have the radial velocity data needed to better constrain the orbit and masses. We will also explore a simultaneous fit to the light curve incorporating both the binary star features and the tidally induced pulsations with \verb"PHOEBE 2.1" (see for e.g., \citealt{2018MNRAS.473.5165H}).

\section{SED fitting}

We fit the spectral energy distribution (SED) of MACHO 80.7443.1718 using the 15 photometric measurements spanning $3.6\mu$m thorough U band given in Table \ref{tab:sedfits}
using DUSTY \citep{1997MNRAS.287..799I,2001MNRAS.327..403E} inside a Markov Chain Monte Carlo wrapper \citep{2015MNRAS.452.2195A}. We assumed foreground extinction due to $R_V=3.1$
dust \citep{1989ApJ...345..245C} and used \citet{2004astro.ph..5087C} model atmospheres for the star. We assume the source is in the LMC at a distance of $d_{\rm LMC}=50 \, \rm kpc$ \citep{2013Natur.495...76P}.   Even when assuming minimum luminosity uncertainties of 10\% for each band, the fits have $\chi^2/N_{dof} \simeq 6$ at fixed $T_*$. While this is adequate for determining the luminosity and extinction at fixed temperature, they are not reliable for determining a temperature (especially since they all lie on the Rayleigh-Jeans side of the SED).  The spectroscopic type, B0, indicates a temperature of 
$T_* \simeq 25,000$~K, and for this temperature $\log (L_*/L_\odot) = 5.55 \pm 0.02$
with $E(B-V) \simeq 0.47 \pm 0.02$ mag.  Based on their photometry, \citet{2002ApJS..141...81M} suggest
a higher temperature of $T_* \simeq 39,000$~K, which drives the luminosity
and extinction up to $\log (L_*/L_\odot) = 6.09 \pm 0.02$ and $E(B-V) \simeq 0.55 \pm 0.02$ mag.
We view the spectroscopic temperature as being more reliable, but our
general conclusions depend weakly on the adopted stellar temperature.

\begin{table}
	\centering
	\caption{Photometry used in the SED fits}
	\label{tab:sedfits}
\begin{tabular}{rrrr}
		\hline
		 Magnitude & $\sigma$ & Filter & Reference\\
		\hline
12.411 & 0.033 &  [3.6] & \citet{2006AJ....132.2268M} \\
12.300 & 0.030 &  [4.5] & \citet{2006AJ....132.2268M} \\
13.255 & 0.007 &  I &  \citet{cioni} \\
12.978 & 0.023 & J & \citet{cioni} \\  
13.56 & 0.10 & V  & \citet{2002ApJS..141...81M} \\
 13.67 & 0.10 & B  & \citet{2002ApJS..141...81M} \\
 12.83 & 0.10 & U   &\citet{2002ApJS..141...81M} \\
 13.43 & 0.10 & R  & \citet{2002ApJS..141...81M} \\
 13.020 & 0.022 &  J & \citet{2003yCat.2246....0C} \\
 12.833 & 0.022 &  H & \citet{2003yCat.2246....0C} \\
 12.734 & 0.030 & $K_s$ & \citet{2003yCat.2246....0C} \\
  12.630 & 0.038 & U & \citet{2004AJ....128.1606Z} \\
  13.617 & 0.106 & B & \citet{2004AJ....128.1606Z} \\
  13.608 & 0.262 & V & \citet{2004AJ....128.1606Z} \\
  13.283 & 0.079 & I & \citet{2004AJ....128.1606Z} \\

\hline
\end{tabular}
\end{table}

The results of the SED fits confirm that the source lies in the LMC.  The estimated Galactic extinction towards the source is only  $E(B-V){\simeq}0.06$ mag \citep{2011ApJ...737..103S}, while the fits require $E(B-V) \simeq 0.5$ mag and the LMC is the only likely source of the additional extinction.  If we tried to make the star a $25,000$~K main sequence star with $L \simeq 10^{3.5}L_\odot$, it would lie at a distance of 5~kpc where such young, massive stars should not exist.  

For the $T_*=25,000$~K SED models, the stellar radius is $R_* \simeq 32 R_\odot$.
The implied mass is trickier because it depends on the extent to which it is 
possible to have stripped mass from the star while maintaining a B0 spectral type.
In the PARSEC \citep{2014MNRAS.445.4287T} models, stars with $T_* \simeq 25,000$~K and $L_* \simeq 10^{5.55}L_\odot$
have $M_* \sim 30M_\odot$ and are starting to evolve across the Hertzsprung gap. The implied mass increases if we assume the higher temperature of \citet{2002ApJS..141...81M}.

\begin{figure}
	\includegraphics[width=0.5\textwidth]{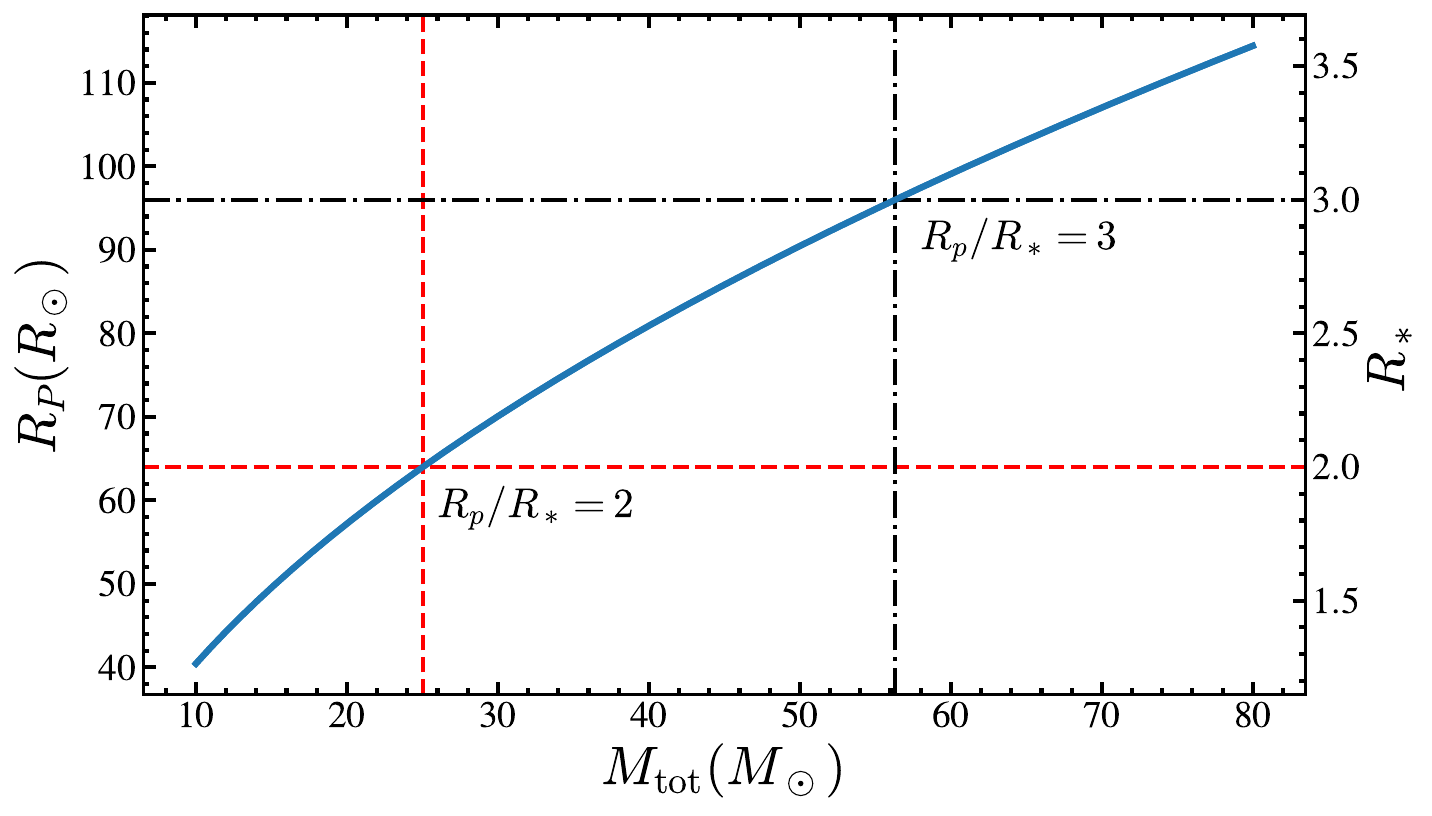}
    \caption{The total mass of the system against the periastron distance. The points at which ${ R_p /R_* }=2$ and ${ R_p /R_* }=3$ are shown as red and black dashed lines respectively.}
    \label{fig:fig4}
\end{figure}

If we combine the stellar radius, the orbital period, and the estimated eccentricity, we can see why the variability amplitudes are so high.
The period and Kepler's third law imply that the
orbital semi-major axis is
\begin{equation}
     a = 93 \left( { M_* + M_c \over 10 M_\odot } \right)^{1/2} R_\odot
\end{equation} 
where $M_* \simeq 30 M_\odot$ is the mass of the star and $M_c$ is the mass of the unobserved companion.
If we assume $e=0.565$ from the fits in section $\S3$ to the ellipsoidal distortions, then the
pericentric radius $R_p = a(1-e)$ in terms of the stellar radius $R_* \simeq 32 R_\odot$ is
\begin{equation}
    { R_p \over R_* } \simeq 1.26 \left( { M_* + M_c \over 10 M_\odot } \right)^{1/2},
\end{equation}
so having $R_p= 2R_*$ implies $M_*+M_c \simeq 25 M_\odot$, and to reach $R_p = 3 R_*$ implies $M_*+M_c \simeq 56 M_\odot$ (Figure \ref{fig:fig4}).   Since
the observed luminosity implies that the visible star is massive,
it appears that the unobserved secondary must also be a massive object
unless the pericentric approach distance is remarkably small.

There is no evidence of accretion (e.g., there seems to be no associated X-ray source in the ROSAT PSPC catalog of X-ray sources in the LMC \citep{1999A&AS..139..277H} and we do not find any \textit{Chandra}/XMM-Newton data), which probably requires that the pericenter lies outside the Roche limit.  This essentially requires that the mass ratio $q=M_c/M_*<1$, since placing the pericenter at the Roche limit for $q=1$ implies $R_p/R_* \simeq 2.6$ and $M_*+M_c>40 M_\odot$.  Assuming the companion is not a black hole, this is consistent with the absence of evidence for emission from the companion star in the SED fits.  If we use the \citet{1985ApJ...295..143M} estimate for the amplitude of the ellipsoidal variability using the pericentric distance for the orbital radius, it is difficult to get amplitudes above $0.1-0.2$\,mag, consistent with our attempts to model the system with \verb"PHOEBE 2.1", without the pericenter lying inside the Roche radius.   However, this is an extrapolation of the \citet{1985ApJ...295..143M} models for both the amplitude and the orbit, so we simply take this as further qualitative evidence that the pericentric radius is $R_p \simeq 2 R_*$.

\begin{table*}
	\centering
	\caption{Pulsation frequencies for MACHO 80.7443.1718, phased to periastron. The errors in frequency, amplitude and phase are calculated through a Monte Carlo analysis.}
	\label{tab:orb}
    \begin{threeparttable}
\begin{tabular}{rrrrrrr}
		\hline
		Data & Frequency $\nu$ ($d^{-1}$) & Orbital Harmonic & $\nu/\nu_{\rm orb}$ & Amplitude (ppt) & Phase & SNR\\
		\hline
		ASAS-SN & & &\\
		& $0.24110 \pm 0.01200$ & 8 & 0.990 & $8.2 \pm 4.2$ & $0.457 \pm 0.252$ & 3.5\\
		& $0.76142 \pm 0.00695$ & 25 & 1.000 & $7.4 \pm 4.5$ & $0.120 \pm 0.225$ & 3.2\\
		\hline
		TESS & & \\
		& $0.21638 \pm 0.0004$ & 7 & 1.015 & $11.1 \pm 0.6$ & $0.180 \pm 0.008$ & 2.7\\
		& $0.76189 \pm 0.0006$ & 25 & 1.000 & $14.9 \pm 0.6$ & $0.283 \pm 0.006$& 4.7\\	
\hline
\end{tabular}

  \end{threeparttable}      
\end{table*}

\section{Tidally Excited Oscillations}

With an approximate model for the tidal distortions, we can subtract the effect of the impulsive forcing and search for tidally excited oscillations (TEOs). TEOs occur at integer multiples of the orbital frequency, thus we carefully consider the orbital harmonics in the FFT spectrum. We calculated the Fast Fourier Transform (FFT) of the residuals using the \verb"Period04" software package \citep{2005CoAst.146...53L} and kept only harmonics with signal-to-noise ratios ($\rm SNRs$) $>2$ for further study. The frequencies were optimized to reduce the light curve residuals. 

We also repeated this analysis for the TESS residuals. The combined TESS light curve shows significant variations outside of periastron with good SNR when compared to the ASAS-SN light curve. In order to reduce the impact of the tidal distortions on this calculation for both the ASAS-SN and TESS data, we only select the epochs with phases in the range [0.25,0.85] (see Figure\ \ref{fig:fig1}). Due to the time-sampling properties and the limited baseline of the TESS data, the FWHM of the peaks in the FFT power spectrum differ (Figure \ref{fig:fig4}).

\begin{figure}
	\includegraphics[width=0.48\textwidth]{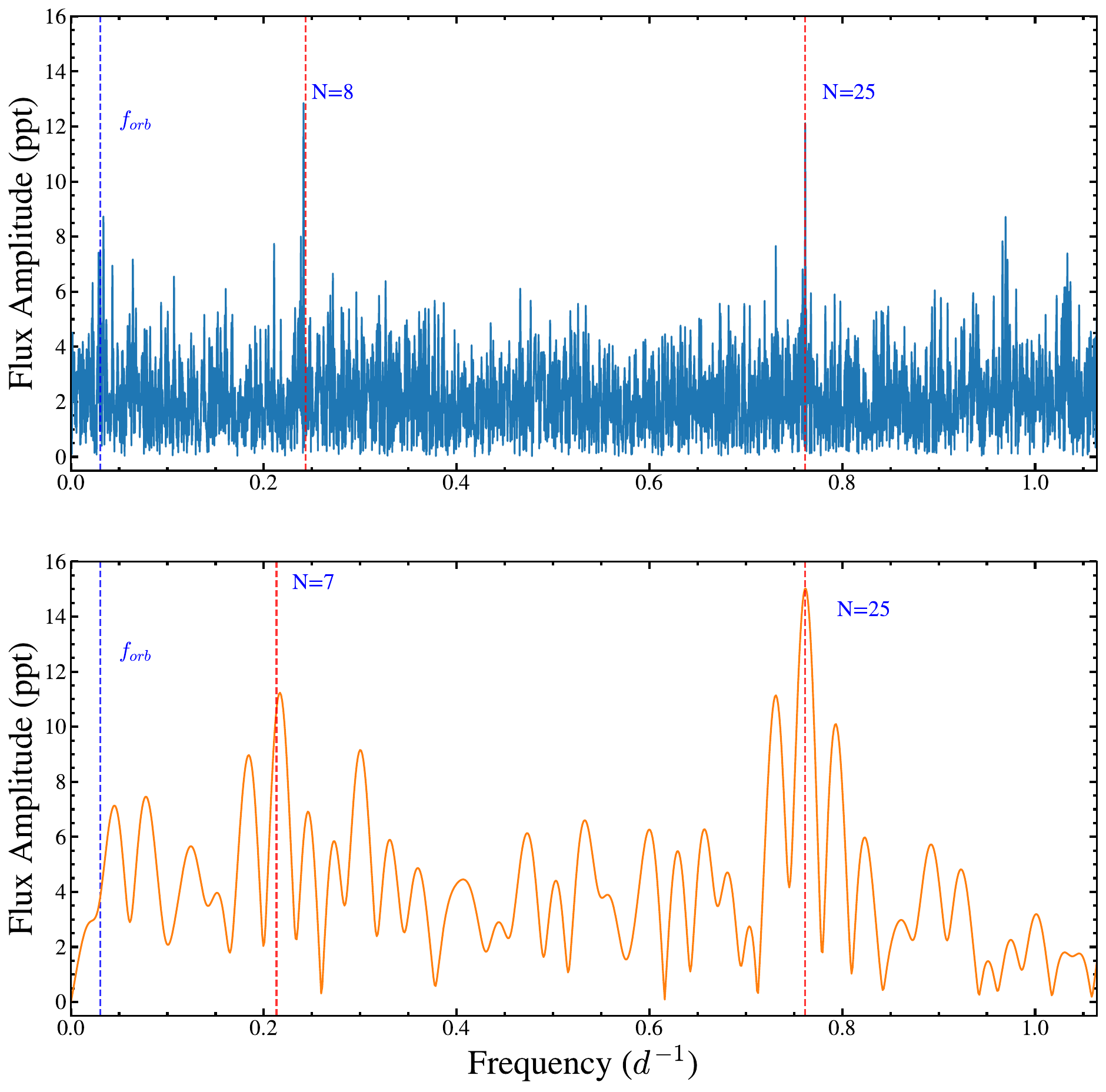}
    \caption{FFT spectrum of the residuals after subtracting the best-fit tidal distortion model from the ASAS-SN V-band data (top) and TESS data (bottom).}
    \label{fig:fig5}
\end{figure}

Figure \ref{fig:fig5} illustrates the FFT power spectrum for the residuals after fitting the best-fit tidal distortion model to the ASAS-SN and TESS data. The significant orbital harmonics are highlighted in red. In the FFT spectrum for the ASAS-SN residuals, we note significant peaks close to $f{\sim}1 \,d^{-1}$. These are likely caused by aliasing and are absent in the TESS spectrum. The TEOs with $\rm SNR>2$ are summarized in Table \ref{tab:orb}. We calculated the uncertainty in the frequencies, amplitudes and phases, 
using the Monte Carlo simulation in \verb"Period04".

The orbital harmonic corresponding to the $N=25$ mode ($P=1.312$ d) was recovered from both the ASAS-SN and TESS data with $\rm SNR>3$. Furthermore, we recover a TEO for the $N=8$ mode in ASAS-SN data and a TEO corresponding to the $N=7$ mode in TESS. The final multi-sector TESS light curve will provide a significantly better characterization of the TEOs.

\section{Conclusions}

We discovered that the variable star MACHO 80.7443.1718 is actually the highest amplitude heartbeat star discovered to date rather than an eclipsing binary. Using both the ASAS-SN and the TESS light curves, we find that
MACHO 80.7443.1718 displays extreme flux variations for a variable of its kind, with maximal variations of ${\sim}40\%$ at periastron and variations of ${\sim}10\%$ due to tidally induced pulsations outside periastron. We fit an analytical model to the light curve to account for the variations caused by tidal distortions and estimated the orbital parameters for this system to be $\omega=-61.4\degree$, $i=44.8\degree$ and $e=0.566$. A more complete model of the system incorporating radial velocity information should improve our constraints of the orbital parameters. The star appears to have tidally induced pulsations at the $N=7/8, \, \rm and\, 25$ harmonics of the orbital period.

MACHO 80.7443.1718 is quite unlike any other heartbeat system discovered --- it is both 
massive and extremely variable for its type. The identification of this source in ASAS-SN and its further characterization using data from the TESS satellite highlights the excellent synergy between these two projects. ASAS-SN is a long baseline survey and provides all-sky light curves that are well suited to study long term variability, whereas TESS light curves are more precise and sampled at a more rapid cadence even though they have a shorter baseline than ASAS-SN. The combination of data from these two surveys will advance the study of variability across the whole sky.

For a more complete characterization of this fascinating system, a radial-velocity follow-up campaign is necessary. These massive heartbeat stars should advance our understanding of the intricacies of stellar evolution and mergers in binary star systems. Furthermore, the tidally induced pulsations in these massive heartbeat systems also probe stellar structure and test theories of dynamical tides. 

\section*{Acknowledgements}
We thank the anonymous referee for their useful comments.
We thank the Las Cumbres Observatory and its staff for its continuing support of the ASAS-SN project.
We thank Jim Fuller for useful comments.

ASAS-SN is supported by the Gordon and Betty Moore Foundation through grant GBMF5490 to the Ohio State University and NSF grant AST-1515927. Development of ASAS-SN has been supported by NSF grant AST-0908816, the Mt. Cuba Astronomical Foundation, the Center for Cosmology and AstroParticle Physics at the Ohio State University, the Chinese Academy of Sciences South America Center for Astronomy (CASSACA), the Villum Foundation, and George Skestos. This work is supported in part by Scialog  Scholar grant 24216 from the Research Corporation. CSK is supported by NSF grants AST-1515876, AST-1515927 and AST-181440. TAT acknowledges support from a Simons Foundation Fellowship and from an IBM Einstein Fellowship from the Institute for Advanced Study, Princeton. 

This paper includes data collected by the TESS mission, which are publicly available from the Mikulski Archive for Space Telescopes (MAST). Funding for the TESS mission is provided by NASA's Science Mission directorate.

This work has made use of data from the European Space Agency (ESA)
mission {\it Gaia} (\url{https://www.cosmos.esa.int/gaia}), processed by
the {\it Gaia} Data Processing and Analysis Consortium (DPAC,
\url{https://www.cosmos.esa.int/web/gaia/dpac/consortium}). 
This paper utilizes public domain data obtained by the MACHO Project,
and has also made use of the VizieR catalogue access tool, CDS, Strasbourg, France. 
This research was made possible through the use of the AAVSO Photometric 
All-Sky Survey (APASS), funded by the Robert Martin Ayers Sciences Fund. 

% Sometimes people leave comments in their papers and other people 
% look for these comments in the version posted on astro-ph, 
% so this is for you, you sickos :)
% ASAS-SN rules!
% Do not remove - KZS

%%%%%%%%%%%%%%%%%%%%%%%%%%%%%%%%%%%%%%%%%%%%%%%%%%



%\bibliographystyle{mnras}

\begin{thebibliography}{99}
\bibitem[Adams \& Kochanek(2015)]{2015MNRAS.452.2195A} Adams, S.~M., \& Kochanek, C.~S.\ 2015, \mnras, 452, 2195
\bibitem[Alard(2000)]{2000A&AS..144..363A} Alard, C.\ 2000, \aaps, 144, 363 
\bibitem[Alard \& Lupton(1998)]{1998ApJ...503..325A} Alard, C., \& Lupton, R.~H.\ 1998, \apj, 503, 325 
\bibitem[Alcock et al.(1997)]{1997ApJ...486..697A} Alcock, C., Allsman, R.~A., Alves, D., et al.\ 1997, \apj, 486, 697 
\bibitem[Bhatti et al.(2018)]{astrob} Bhatti, W., Bouma, L.~G., Wallace, J., et al.\ 2018, astrobase, v0.3.8, Zenodo, http://doi.org/10.5281/zenodo.1185231
\bibitem[Burkart et al.(2012)]{2012MNRAS.421..983B} Burkart, J., Quataert, E., Arras, P., et al.\ 2012, \mnras, 421, 983.
\bibitem[Cardelli et al.(1989)]{1989ApJ...345..245C} Cardelli, J.~A., Clayton, G.~C., \& Mathis, J.~S.\ 1989, \apj, 345, 245 
\bibitem[Castelli \& Kurucz(2004)]{2004astro.ph..5087C} Castelli, F., \& Kurucz, R.~L.\ 2004, arXiv:astro-ph/0405087 
\bibitem[Cioni et al.(2000)]{cioni} Cioni, M.-R.~L., van der Marel, R.~P., Loup, C., \& Habing, H.~J.\ 2000, \aap, 359, 601 
\bibitem[Clarke(2002)]{2002A&A...386..763C} Clarke, D.\ 2002, \aap, 386, 763 
\bibitem[Cutri et al.(2003)]{2003yCat.2246....0C} Cutri, R.~M., Skrutskie, M.~F., van Dyk, S., et al.\ 2003, VizieR Online Data Catalog, 2246,
\bibitem[Elitzur \& Ivezi{\'c}(2001)]{2001MNRAS.327..403E} Elitzur, M., \& Ivezi{\'c}, {\v Z}.\ 2001, \mnras, 327, 403 
\bibitem[Fuller, \& Lai(2012)]{2012MNRAS.420.3126F} Fuller, J., \& Lai, D.\ 2012, \mnras, 420, 3126.
\bibitem[Fuller(2017)]{2017MNRAS.472.1538F} Fuller, J.\ 2017, \mnras, 472, 1538.
\bibitem[Foreman-Mackey et al.(2013)]{2013PASP..125..306F} Foreman-Mackey, D., Hogg, D.~W., Lang, D., et al.\ 2013, \pasp, 125, 306
\bibitem[Gaia Collaboration et al.(2018)]{2018arXiv180409365G} Gaia Collaboration, Brown, A.~G.~A., Vallenari, A., et al.\ 2018, arXiv:1804.09365 
\bibitem[Garmany et al.(1994)]{1994AJ....108.1256G} Garmany, C.~D., Massey, P., \& Parker, J.~W.\ 1994, \aj, 108, 1256.
\bibitem[Haberl \& Pietsch(1999)]{1999A&AS..139..277H} Haberl, F., \& Pietsch, W.\ 1999, \aaps, 139, 277 
\bibitem[Hambleton et al.(2018)]{2018MNRAS.473.5165H} Hambleton, K., Fuller, J., Thompson, S., et al.\ 2018, \mnras, 473, 5165.
\bibitem[Henden et al.(2015)]{2015AAS...22533616H} Henden, A.~A., Levine, S., Terrell, D., \& Welch, D.~L.\ 2015, American Astronomical Society Meeting Abstracts \#225, 225, 336.16
\bibitem[Horvat et al.(2018)]{2018ApJS..237...26H} Horvat, M., Conroy, K.~E., Pablo, H., et al.\ 2018, \apjs, 237, 26
\bibitem[Ivezic \& Elitzur(1997)]{1997MNRAS.287..799I} Ivezic, Z., \& Elitzur, M.\ 1997, \mnras, 287, 799 
\bibitem[Jayasinghe et al.(2018a)]{2018MNRAS.477.3145J} Jayasinghe, T., Kochanek, C.~S., Stanek, K.~Z., et al.\ 2018, \mnras, 477, 3145 
\bibitem[Jayasinghe et al.(2019a)]{2018arXiv180907329J} Jayasinghe, T., Stanek, K.~Z., Kochanek, C.~S., et al.\ 2019, \mnras, 486, 1907
\bibitem[\protect\citeauthoryear{Jayasinghe et al.}{2019b}]{2019MNRAS.485..961J} Jayasinghe T., et al., 2019, MNRAS, 485, 961
\bibitem[Kirk et al.(2016)]{2016AJ....151...68K} Kirk, B., Conroy, K., Pr{\v s}a, A., et al.\ 2016, \aj, 151, 68 
\bibitem[Kochanek et al.(2017)]{2017PASP..129j4502K} Kochanek, C.~S., Shappee, B.~J., Stanek, K.~Z., et al.\ 2017, \pasp, 129, 104502 
\bibitem[Kov{\'a}cs et al.(2002)]{2002A&A...391..369K} Kov{\'a}cs, G., Zucker, S., \& Mazeh, T.\ 2002, \aap, 391, 369 
\bibitem[Kumar et al.(1995)]{1995ApJ...449..294K} Kumar, P., Ao, C.~O., \& Quataert, E.~J.\ 1995, \apj, 449, 294 
\bibitem[Lafler \& Kinman(1965)]{1965ApJS...11..216L} Lafler, J., \& Kinman, T.~D.\ 1965, \apjs, 11, 216 
\bibitem[Lenz \& Breger(2005)]{2005CoAst.146...53L} Lenz, P., \& Breger, M.\ 2005, Communications in Asteroseismology, 146, 53.
\bibitem[Lindegren et al.(2018)]{2018A&A...616A...2L} Lindegren, L., Hern{\'a}ndez, J., Bombrun, A., et al.\ 2018, \aap, 616, A2.
\bibitem[Massey(2002)]{2002ApJS..141...81M} Massey, P.\ 2002, The Astrophysical Journal Supplement Series, 141, 81.
\bibitem[Maceroni et al.(2009)]{2009A&A...508.1375M} Maceroni, C., Montalb{\'a}n, J., Michel, E., et al.\ 2009, \aap, 508, 1375.
\bibitem[Meixner et al.(2006)]{2006AJ....132.2268M} Meixner, M., Gordon, K.~D., Indebetouw, R., et al.\ 2006, \aj, 132, 2268.
\bibitem[Morris(1985)]{1985ApJ...295..143M} Morris, S.~L.\ 1985, \apj, 295, 143 
\bibitem[Murray \& Correia(2010)]{2010exop.book...15M} Murray, C.~D., \& Correia, A.~C.~M.\ 2010, Exoplanets, 15.
\bibitem[Nieva \& Przybilla(2014)]{2014A&A...566A...7N} Nieva, M.-F., \& Przybilla, N.\ 2014, \aap, 566, A7.
\bibitem[Pablo et al.(2017)]{2017MNRAS.467.2494P} Pablo, H., Richardson, N.~D., Fuller, J., et al.\ 2017, \mnras, 467, 2494.
\bibitem[Pietrzy{\'n}ski et al.(2013)]{2013Natur.495...76P} Pietrzy{\'n}ski, G., Graczyk, D., Gieren, W., et al.\ 2013, \nat, 495, 76 
\bibitem[Pedregosa et al.(2012)]{2012arXiv1201.0490P} Pedregosa, F., Varoquaux, G., Gramfort, A., et al.\ 2012, arXiv:1201.0490 
\bibitem[Pr{\v{s}}a et al.(2016)]{2016ApJS..227...29P} Pr{\v{s}}a, A., Conroy, K.~E., Horvat, M., et al.\ 2016, \apjs, 227, 29
\bibitem[Ricker et al.(2015)]{2015JATIS...1a4003R} Ricker, G.~R., Winn, J.~N., Vanderspek, R., et al.\ 2015, Journal of Astronomical Telescopes, Instruments, and Systems, 1, 014003
\bibitem[Scargle(1982)]{1982ApJ...263..835S} Scargle, J.~D.\ 1982, \apj, 263, 835
\bibitem[Schlafly \& Finkbeiner(2011)]{2011ApJ...737..103S} Schlafly, E.~F., \& Finkbeiner, D.~P.\ 2011, \apj, 737, 103 
\bibitem[Schwarzenberg-Czerny(1996)]{1996ApJ...460L.107S} Schwarzenberg-Czerny, A.\ 1996, \apjl, 460, L107 
\bibitem[Shappee et al.(2014)]{2014ApJ...788...48S} Shappee, B.~J., Prieto, J.~L., Grupe, D., et al.\ 2014, \apj, 788, 48
\bibitem[Shporer et al.(2016)]{2016ApJ...829...34S} Shporer, A., Fuller, J., Isaacson, H., et al.\ 2016, \apj, 829, 34.
\bibitem[Tang et al.(2014)]{2014MNRAS.445.4287T} Tang, J., Bressan, A., Rosenfield, P., et al.\ 2014, \mnras, 445, 4287 
\bibitem[Thompson et al.(2012)]{2012ApJ...753...86T} Thompson, S.~E., Everett, M., Mullally, F., et al.\ 2012, \apj, 753, 86.
\bibitem[Watson et al.(2006)]{2006SASS...25...47W} Watson, C.~L., Henden, A.~A., \& Price, A.\ 2006, Society for Astronomical Sciences Annual Symposium, 25, 47
\bibitem[Welsh et al.(2011)]{2011ApJS..197....4W} Welsh, W.~F., Orosz, J.~A., Aerts, C., et al.\ 2011, The Astrophysical Journal Supplement Series, 197, 4.
\bibitem[Zechmeister \& K{\"u}rster(2009)]{2009A&A...496..577Z} Zechmeister, M., \& K{\"u}rster, M.\ 2009, \aap, 496, 577
\bibitem[Zaritsky et al.(2004)]{2004AJ....128.1606Z} Zaritsky, D., Harris, J., Thompson, I.~B., \& Grebel, E.~K.\ 2004, \aj, 128, 1606 

\end{thebibliography}

\vspace{1cm}


%%%%%%%%%%%%%%%%%%%%%%%%%%%%%%%%%%%%%%%%%%%%%%%%%%

%%%%%%%%%%%%%%%%% APPENDICES %%%%%%%%%%%%%%%%%%%%%

%%%%%%%%%%%%%%%%%%%%%%%%%%%%%%%%%%%%%%%%%%%%%%%%%%

% Don't change these lines
\bsp	% typesetting comment
\label{lastpage}
\end{document}